# Pseudogap and charge density waves in two dimensions

S. V. Borisenko[1], A. A. Kordyuk[1,2], A. N. Yaresko[3], V. B. Zabolotnyy[1], D. S. Inosov[1], R. Schuster[1], B. Büchner[1], R. Weber[4], R. Follath[4], L. Patthey[5], H. Berger[6]

[1]*Leibniz-Institute for Solid State Research, IFW-Dresden, D-01171, Dresden, Germany*

[2]*Institute of Metal Physics, 03142 Kyiv, Ukraine*

[3]*Max-Planck-Institute for the Physics of Complex Systems, Dresden, Germany*

[4]*BESSY, Berlin, Germany*

[5]*Swiss Light Source, Paul Scherrer Institut, CH-5234 Villigen, Switzerland*

[6]*Institute of Physics of Complex Matter, EPFL, 1015 Lausanne, Switzerland*

**An interaction between electrons and lattice vibrations (phonons) results in two fundamental quantum phenomena in solids: in three dimensions it can turn a metal into a superconductor whereas in one dimension it can turn a metal into an insulator[1,2,3]. In two dimensions (2D) both superconductivity and charge-density waves (CDW)[4,5] are believed to be anomalous. In superconducting cuprates, critical transition temperatures are unusually high and the energy gap may stay unclosed even above these temperatures (pseudogap). In CDW-bearing dichalcogenides the resistivity below the transition can decrease with temperature even faster than in the normal phase[6,7] and a basic prerequisite for the CDW, the favourable nesting conditions (when some sections of the Fermi surface appear shifted by the same vector), seems to be absent[8,9,10]. Notwithstanding the existence of alternatives[11,12,13,14,15] to conventional theories[1,2,3], both phenomena in 2D still remain the most fascinating puzzles in condensed matter physics. Using the latest developments in high-resolution angle-resolved photoemission spectroscopy (ARPES) here we show that the normal-state pseudogap also exists in one of the most studied 2D examples, dichalcogenide 2H-TaSe$_2$, and the formation of CDW is driven by a conventional nesting instability, which is masked by the pseudogap. Our findings reconcile and explain a number of unusual, as previously believed, experimental responses as well as disprove many alternative theoretical approaches[12,13,14,15]. The magnitude, character and anisotropy of the 2D-CDW pseudogap are intriguingly similar to those seen in superconducting cuprates.**

Variations of the electron density in a metal are highly unfavourable because of the Coulomb repulsion. Though, some low-dimensional systems such as transition-metal dichalcogenides spontaneously develop a static periodic modulation (known as charge-density wave) of the electron gas below a certain temperature. A typical representative of the transition metal dichalcogenides is a 2H (trigonal prismatic) polytype of $TaSe_2$ which exhibits two CDW phase transitions at accessible temperatures: a second-order one at $T_{NIC}$=122 K from normal to an incommensurate CDW state and a first-order lock-in transition at $T_{ICC}$=90 K from the incommensurate to a 3x3 commensurate CDW phase[16]. In the one-dimensional case, where all points of the Fermi surface (FS) can be connected by the same vector (perfect nesting), the CDW transition occurs when the energy gain due to opening of a gap at the Fermi level exceeds the energy costs to distort the lattice[2,4,17]. In real 2D materials the energy balance is more delicate since all FS points cannot be connected by the same vector (non-perfect nesting) and thus the FS can be gapped only partially. To understand the CDW mechanism in 2H-$TaSe_2$, a detailed knowledge of the low-energy electronic structure is required. We therefore start with an overview of its temperature evolution in Figs. 1 and 2.

The upper panel of Fig. 1 shows the topology of the Fermi surface in the normal state together with the CDW vectors $\vec{q}_n = \frac{2}{3}\Gamma M$ defined by other experiments[5,16]. Contrary to the earlier band structure calculations[8], but in accordance with a recent study[10], the FS consists of single hole-like "barrels" centred at $\Gamma$ and K points, and electron-like "dogbones" around the M-point. The upper row of panels in Fig. 2a shows the corresponding dispersions of the electronic states crossing the Fermi level. FS sheets originate from two bands: one is responsible for the $\Gamma$ and K barrels with a saddle point in between, the other one supports the dogbone with M being another saddle point. The second row of panels in Fig. 2a showing the data below the first phase transition, suggests that the normal ( 290 K ) and incommensurate CDW state ( $T_{ICC}$ < 107 K < $T_{NIC}$ ) dispersions are qualitatively similar, except for the naturally different temperature broadening and weaker crossing #3 (Fig. 2a). This seems to be in agreement with the surprising earlier ARPES results, when virtually no change of the electronic structure has been detected upon entering the incommensurate CDW phase[9,10,18], thus clearly contradicting the concept of the energy gain that should accompany a CDW transition. In contrast, the lock-in transition to the commensurate CDW state at $T_{ICC}$=90 K is much more pronounced (see lower panel of Fig.1). The new folded FS is schematically shown in Fig. 1 as a set of nearly circles around new $\Gamma$"-points and rounded triangles around new K'-points. This topology of the folded FS, though natural, as suggested by the normal state FS (see Supplementary material), has never been detected before. A possible reason could be a rather weak umklapp potential due to small lattice distortion of the order of 0.05 Å (Ref 16, 17), which consequently results in the intensity distribution along the FS that still reminds the one seen in the normal state. Note, that without such an overview of the large portion of the k-space, it is problematic to understand what exactly happens to the electronic structure below 90 K. The energy-momentum intensity distributions in the lower row of panels in Fig. 2a are also considerably modified by the 3x3 folding and show clear signatures of a strong hybridization. Exactly this hybridization, when occurs in the vicinity of the Fermi level, can lower the energy of the system (see right inset to Fig. 3c). This observation of strong changes at $T_{ICC}$ is again paradoxical. According to other experimental techniques, the CDW phase transition with the *critical* energy lowering occurs at $T_{NIC}$=122 K as is clearly seen in e.g. temperature dependences of the specific heat or resistivity[6, 7]. The lock-in transition at $T_{ICC}$=90 K, in contrast, is hardly detectable in the mentioned curves[6, 7] and appears as a small break in the superlattice strength as seen by neutron scattering[16].

More detailed data analysis clarifies the situation. ARPES offers a unique opportunity to find out whether the energy gap opens up at the Fermi level anywhere on the FS by tracking the binding energy of the leading edge of an energy distribution curve (EDC) taken at $k_F$. In Fig. 2b we show EDCs corresponding exactly to $k_F$ for selected high symmetry cuts in the k-space (Fig. 2a). Both panels unambiguously signal the leading edge shift (~15 meV) of the EDC #6. In both cases a clear suppression (not absence) of the spectral weight at the Fermi level is evident also from the corresponding energy-momentum distributions (rightmost panels in Fig. 2a), which is reinforced by an obviously more diffused appearance of the K-barrels on the normal-state FS map (Fig. 1). In a striking analogy with the superconducting cuprates, we thus conclude the presence of a pseudogap in both, normal and incommensurate CDW states of 2H-TaSe$_2$.

In order to understand whether necessary energy gain can come from the pseudogap, we further characterize the pseudogap as a function of temperature and momentum in Fig. 3. The so called maps of gaps[19], the plots of binding energies of the EDC leading edges as a function of k (i.e. not only $k_F$), are shown for the normal and incommensurate CDW states in Fig. 3a,b. These maps are a visual demonstration of the energy lowering of the system and are ideal for the determination of the anisotropy of the gap as they make the analysis of the behaviour of the leading edge in the vicinity of the Fermi surface (shown as dotted lines) possible. While the normal state map (Fig. 3a) reveals an isotropic pseudogap detectable only on the K-barrel, in the CDW state a pseudogap opens, in addition, on the dogbone FS and is anisotropic. As a quantitative measure of the pseudogap we take the difference of the leading edge binding energies of the $k_F$-EDCs of M-dogbone and K-barrel which belong to the M-K cut, as is sketched in the left inset to the Fig. 3c. The sharp reproducible increase of the pseudogap magnitude below ~122 K, which escaped the detection before, is distinctly seen in Fig. 3c. It is now clear that it is the NIC transition ($T_{NIC}$=122K) at which the *critical* lowering of the electronic energy occurs, and not only because of the larger pseudogap on the K-barrel, but also owing to the opening of the anisotropic pseudogap on the dogbone M-centred FS. In the commensurate CDW phase one can no longer characterize the energy gap by the leading edge position because of the interference with the folded bands --- the rounded corners of the new small triangular FS fall right where the normal state K-barrel was located (see Fig. 1). Instead, we plot in Fig. 3c also the values determined as shown in the right inset. The "band-gap" is a direct consequence of the hybridization, distinctly observed below $T_{ICC}$ (Figs. 1, 2). In order to emphasize the existence of a crossover regime where the pseudogap evolves into a band-gap we also plot in a limited T- interval the leading edge positions below $T_{ICC}$ and the band-gap above $T_{ICC}$. Both gaps do not exhibit any anomalies at $T_{ICC}$ monotonically increasing upon cooling deeper in the CDW state.

Presented data already put certain limitations on several alternative theoretical approaches that were stimulated by the earlier experiments. The locations of the saddle points in the momentum-energy space obviously do not support the saddle point nesting scenario[13], proposed as an alternative to the conventional nesting, as both points are far from the Fermi level (280 meV and 330 meV) and are not connected by any of the CDW vectors (the first one is located between Γ- and K-barrels, and the second is the M-point itself) . The pseudogap supported by the K-centred FS stays isotropic (within ~ 2 meV) thus ruling out a six-fold symmetric CDW gap with nodes as suggested in Ref. 14. The value of the band-gap saturates at low temperatures at ~33 meV which is nearly a factor of 5 smaller than the one (~ 150 meV) obtained in the strong-coupling approach[12]. A recent proposal[15] to consider two components of the electronic structure, one of which is gapped and the other one is not, seems to be not supported by the data as well.

What is then responsible for the CDW in two dimensions? In the following we demonstrate that the conventional FS nesting scenario, though modified by the presence of the normal-state pseudogap, is perfectly applicable. From our high-resolution measurements, we have analysed the nesting properties of the FS quantitatively, which made it possible to explain the temperature evolution of the electronic structure of TaSe$_2$ step by step. The quantitative measure of the nesting is the charge susceptibility. Here we approximate the charge susceptibility by the autocorrelation of the FS map[20]. In order to avoid the influence of the matrix elements, for further processing we take the model FS map shown in Fig. 4a, which is an exact copy of the experimental one as far as the locus of $k_F$-points is concerned. In addition, this gives possibility to investigate the FS nesting properties of a *hypothetical* compound, with the electronic structure that yet have not reacted to the nesting instability (i.e. without gaps). The resulting autocorrelation maps as a function of temperature are shown in Fig. 4b, while the corresponding cuts along the ΓM direction -- in Fig. 4d. The sharp peak, seen in the 290 K curve exactly at 2/3 ΓM in Fig. 4d, is the first clear evidence for the nearly perfect nesting in 2D chalcogenides. This is at variance with the previous calculations[17] which have found susceptibility for 2H polytypes to take a broadly humped form. Thus, the Fermi surface shape alone, without taking into account the pseudogap, results in the peak in the charge susceptibility at a wave vector ~2/3ΓM. According to the neutron scattering[16], exactly at this wave vector there is a strong Kohn-like anomaly[4] of the $\Sigma_1$ phonon branch already at 300 K, and the matching phonon softens even more as the transition is approached. We consider therefore the formation of both, normal state pseudogap and anomaly of the $\Sigma_1$ phonon branch as respective reactions of the electronic and lattice subsystems to the instability caused by the strong scattering channel that appears due to the nesting and a presence of the suitable mediating phonon. Such a mutual response can signify a strong electron-phonon interaction in 2H-TaSe$_2$. It is interesting, that despite the favourable conditions, the system does not escape the instability by developing a static commensurate CDW order, presumably because of still too high temperature, which would effectively close not large enough band-gaps. Instead, it opens up a pseudogap. Upon cooling we observe correlated changes of the susceptibility, pseudogap magnitude, and energy of the softened phonon[16]: the peak in the susceptibility splits into two (middle panel in Fig. 4d), the pseudogap slowly closes (as suggested by the EDC shift at 290 K in Fig. 2b and a weak but detectable trend of the gap to decrease in Fig. 3c) , the phonon energy becomes lower. This worsening of the nesting conditions is fundamentally different from the 1D case, where the susceptibility exhibits a peak, which becomes sharper with lowering the temperature[4]. We have found out that such anomalous behaviour of the susceptibility, i.e. the splitting and the size of the peak at ~2/3ΓM, is very sensitive to the shape of the Fermi surface itself, namely to the distance between the dogbone and K-centred FSs (distance, marked 'D' in Fig. 4a). The data plotted in Fig. 4c clearly suggest that the FS itself is temperature dependent explaining the variation of the charge susceptibility. Moreover, a close correlation with the temperature dependence of the pseudogap magnitude (Fig. 3c) implies that the pseudogap itself is directly related to the charge susceptibility. Nevertheless, the transition finally occurs at $T_{NIC}$=122 K , but now only into the incommensurate CDW phase as dictated by the split peak in the susceptibility. Further reduction of the temperature results in a reversed modification of the obviously correlated quantities: pseudogap opens up more, peaks in susceptibility move towards each other (right panel in Fig. 4d), which agrees well with the dynamics of the superlattice peaks seen by neutrons, and phonon energy starts to increase again[16]. In such a manner the system arrives at the transition into the commensurate CDW state at 90 K.

Our findings provide a natural explanation not only for the neutron scattering experiments[16]. Nearly linear in-plane resistivity in the normal state[6, 7] is similar to the resistivity of an optimally doped cuprate superconductor[21] and shows a typical behaviour of a pseudo-gapped metal. Below 122 K the slope is increasing in close correspondence to the larger pseudogap, and below 90 K the resistivity resembles the one of a normal metal. Optical measurements on the same single crystals[7] have indirectly suggested the presence of a pseudogap already at 300 K as well, though with somewhat different energy scale. We also notice an excellent agreement with the Hall coefficient measurements[22]. According to Ref.22, the Hall coefficient starts to decrease sharply from its positive value below ~120 K, and changes sign at 90 K. The positive value in the normal state is explained by the larger volume of the Γ- and K-centred hole-like barrels in comparison with the electron-like dogbones around the M-points. The sharp increase of the pseudogap at 122 K results in the reduction of the number of charge carriers and finally in the commensurate phase (below 90 K), the area enclosed by the electron-like circular FS around new Γ'-points is clearly larger than the two (per new BZ) hole-like triangular FS, which leads to the negative sign of the Hall coefficient.

It was suggested before[23] that the HTSCs and 2D chalcogenides have similar phase diagrams and that the pseudogap regime in cuprates is very similar to the CDW regime of chalcogenides. Now, after observation of the pseudogap in $TaSe_2$ we can directly compare both pseudogaps. The energy-momentum distribution of the photoemission intensity near $k_F$ of the K-barrel (crossing #6, Fig. 2a) is very similar to the one measured for the bonding barrel in the underdoped Bi2212-cuprate[24, 25]. In both cases one can still track the dispersion up to the Fermi level, but the spectral weight is significantly suppressed resulting in the shift of the $k_F$-EDC's leading edge. Furthermore, the detected anisotropy of the pseudogap on the dogbone FS is reminiscent of the famous anisotropic behaviour of the pseudogap in cuprates. Finally, the highly inhomogeneous intensity distribution along some of the small triangular FSs and especially along incomplete circles around new Γ' points cannot escape a comparison with the famous Fermi surface 'arcs'. We believe that this series of remarkable similarities to the high-temperature superconducting cuprates calls for more careful comparative studies of the pseudogap phenomenon in these materials.

It is interesting, that unlike in the 1D case where a pseudogap has been reported[26, 27, 28] to be a consequence of fluctuations (potentially able to suppress the transition temperature up to a quarter of the mean-field value[29]), the pseudogap in 2D shows unusual non-monotonic behaviour clearly tracking the temperature evolution of both, the bare susceptibility and phonon spectrum, and thus seems to represent a natural response of the system to a nesting instability.

Acknowledgements: The project was supported, in part, by the DFG under Grant No. KN393/4. We thank R. Hübel for technical support. This work was partially performed at the Swiss Light Source, Paul Scherrer Institut, Villigen, Switzerland.

Correspondence and requests for materials should be addressed to S.V.B. (S. Borisenko@ifw-dresden.de).

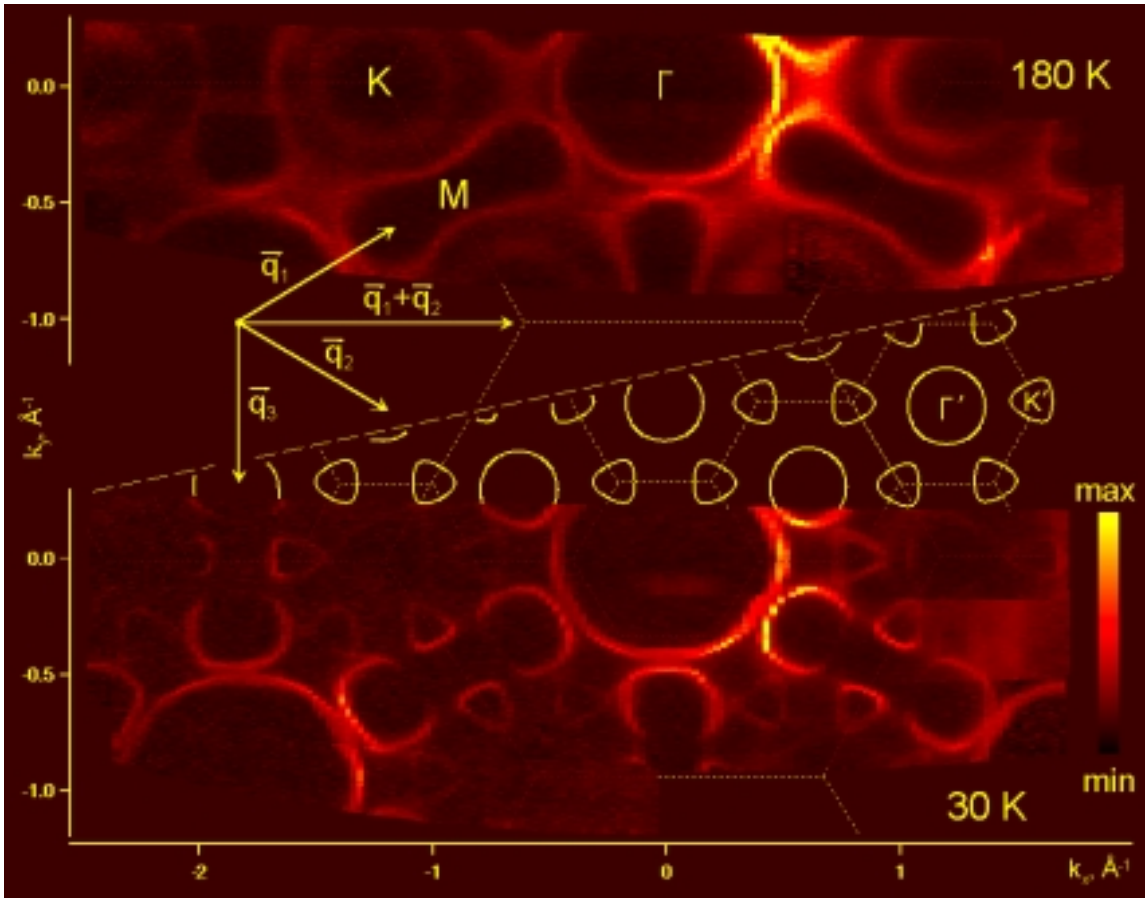

**Figure 1. Fermi surfaces.** Momentum distribution of the photoemission intensity at the Fermi level at 180 K and 30 K. Short dashed lines are the BZ boundaries.

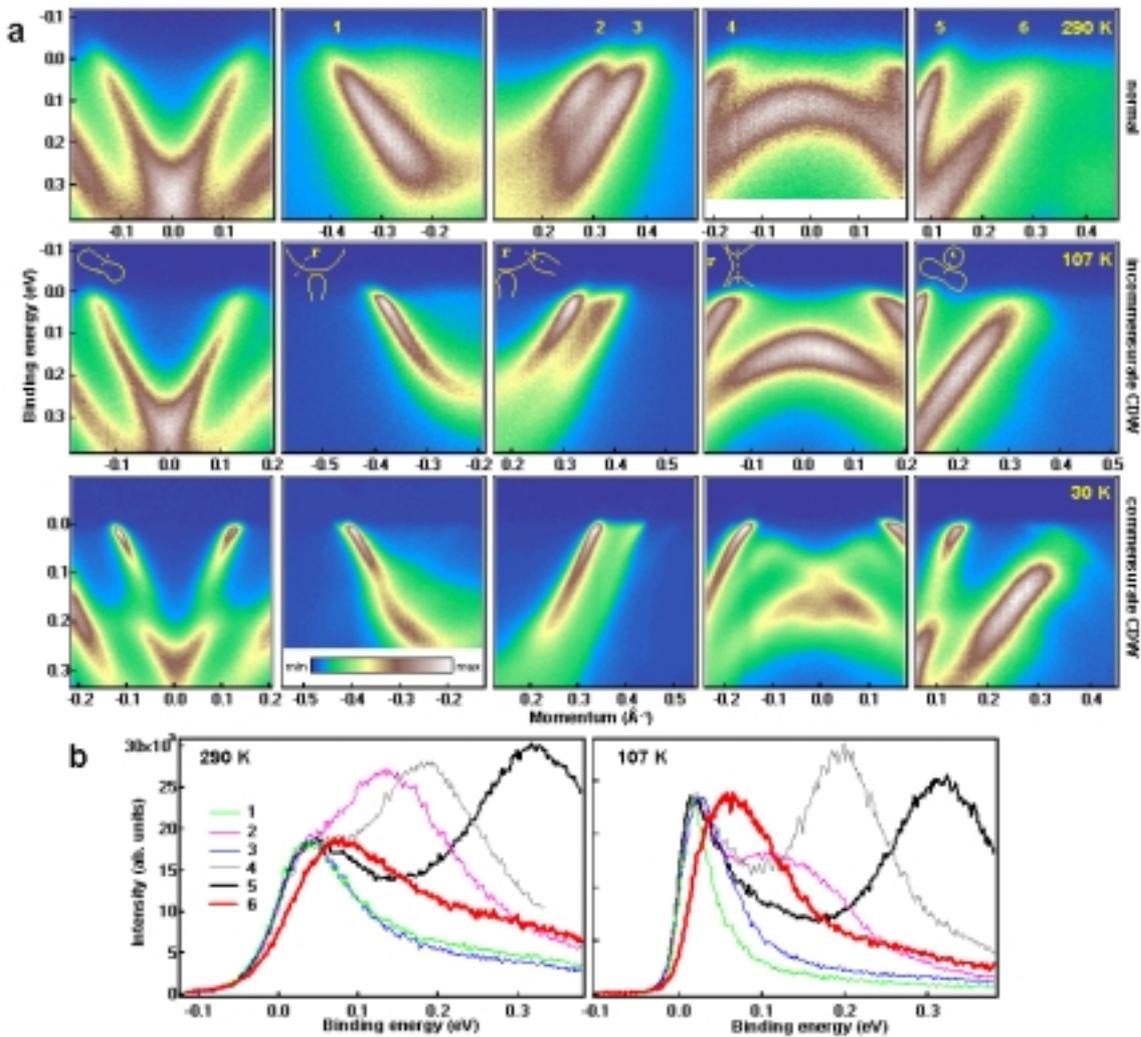

**Figure 2. Electronic structure and leading edge gap.** a) Photoemission intensity as a function of energy and momentum in the normal (upper row of panels), incommensurate CDW (middle row) and commensurate CDW (lower row) states. Sketches of the FS on the middle panels show cuts in momentum space along which the data were taken and are valid for all panels in the same column. Hybridization effects in the lower row are seen as "repulsions" of the bands, which occur when a folded band is supposed to cross the original one, as schematically shown in the right inset to Fig. 3b. Since the spectral weight of the folded band is lower, these effects appear as breaks in the intensity of the original bands. Numbers correspond to the different $k_F$. b) $k_F$ EDCs from the datasets shown in a). Leading edge gap is clearly seen as a shift of the EDC#6 to the higher binding energies at 290 K and 107 K.

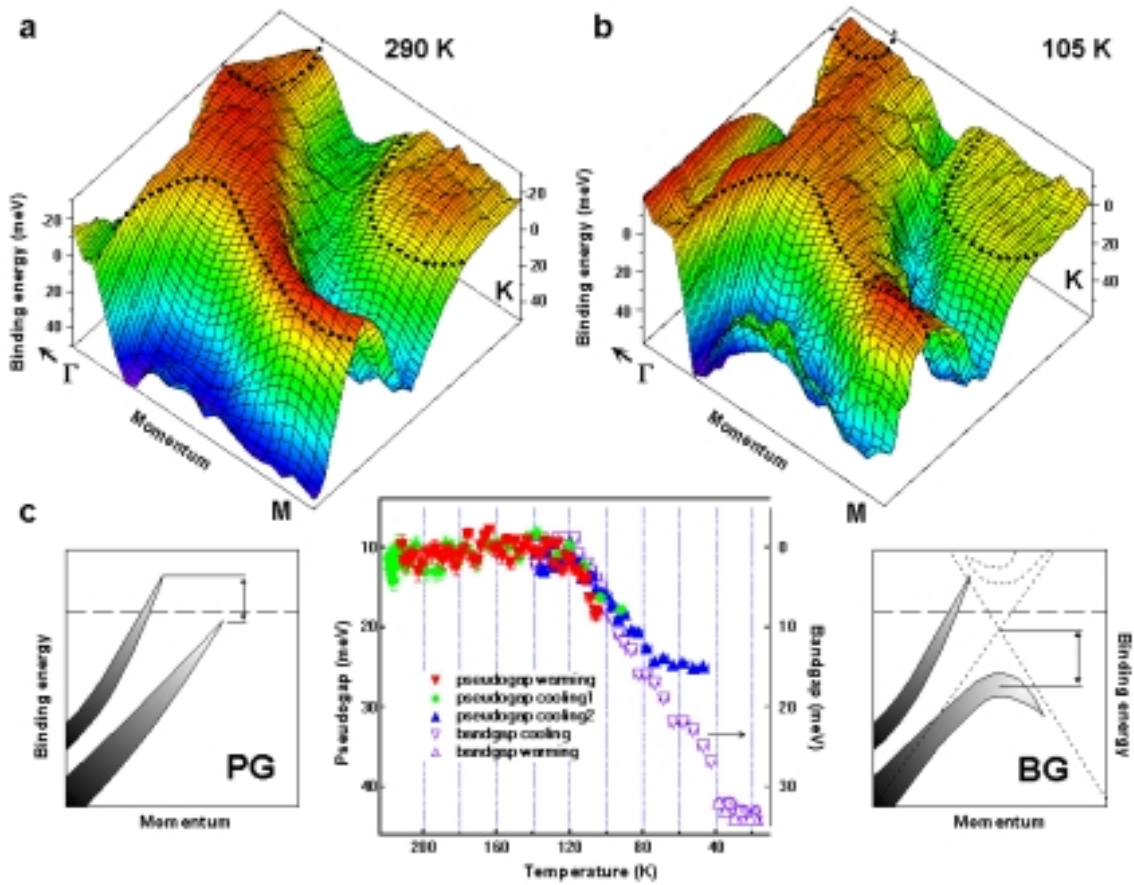

**Figure 3. Momentum and temperature dependence of the pseudogap.** a,b) Binding energies of the leading edges of all EDCs from the momentum range close to the irreducible parts of the M-dogbone and K-barrel FS. Colour scales reproduce the vertical coordinate. Anisotropy of the pseudogap on the M-dogbone FS is seen as a changing colour when going along the dashed line which correspond to FS, i.e. $k_F$ points. c) Difference between the binding energies of the leading edges of the EDC#5 and #6 from Fig. 2b as a function of temperature as shown schematically in the left inset (pseudogap) when cycling the temperature (filled symbols). Shift of the EDC maximum (bandgap) which corresponds to the top of the hybridized band as shown in the right inset with respect to its position at $T_{NIC}$=122 K (open symbols). Note that the given leading edge gap values below 90 K cannot be considered as a measure of the pseudogap because of folding.

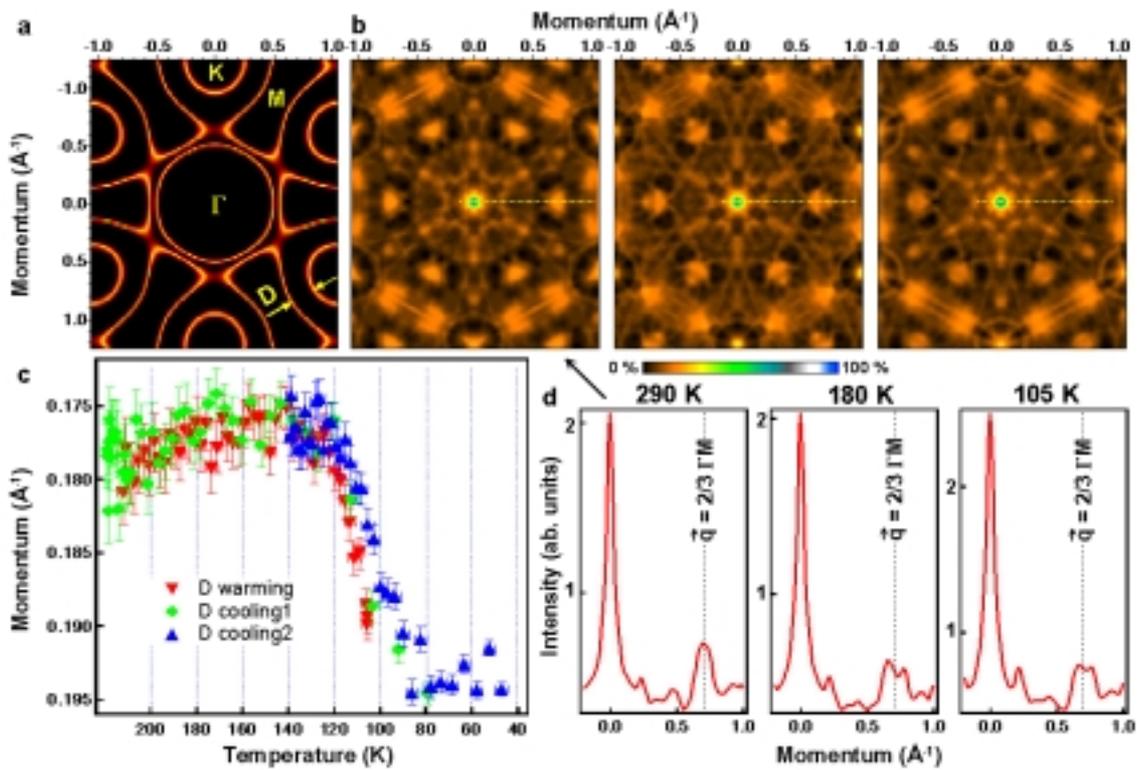

**Figure 4. Nesting properties.** a) A model copy of the FS map from Fig. 1 with the homogeneous intensity distribution along the FS. b) Autocorrelation maps of this model at different temperatures. c) Temperature dependence of the momentum distance between the M-dogbone and K-barrel (T > 90 K). The distance below 90 K is just a difference between corresponding maxima at the FS map. d) Intensity of the maps from b) along the cuts shown by dotted lines.

**Supplementary information.**

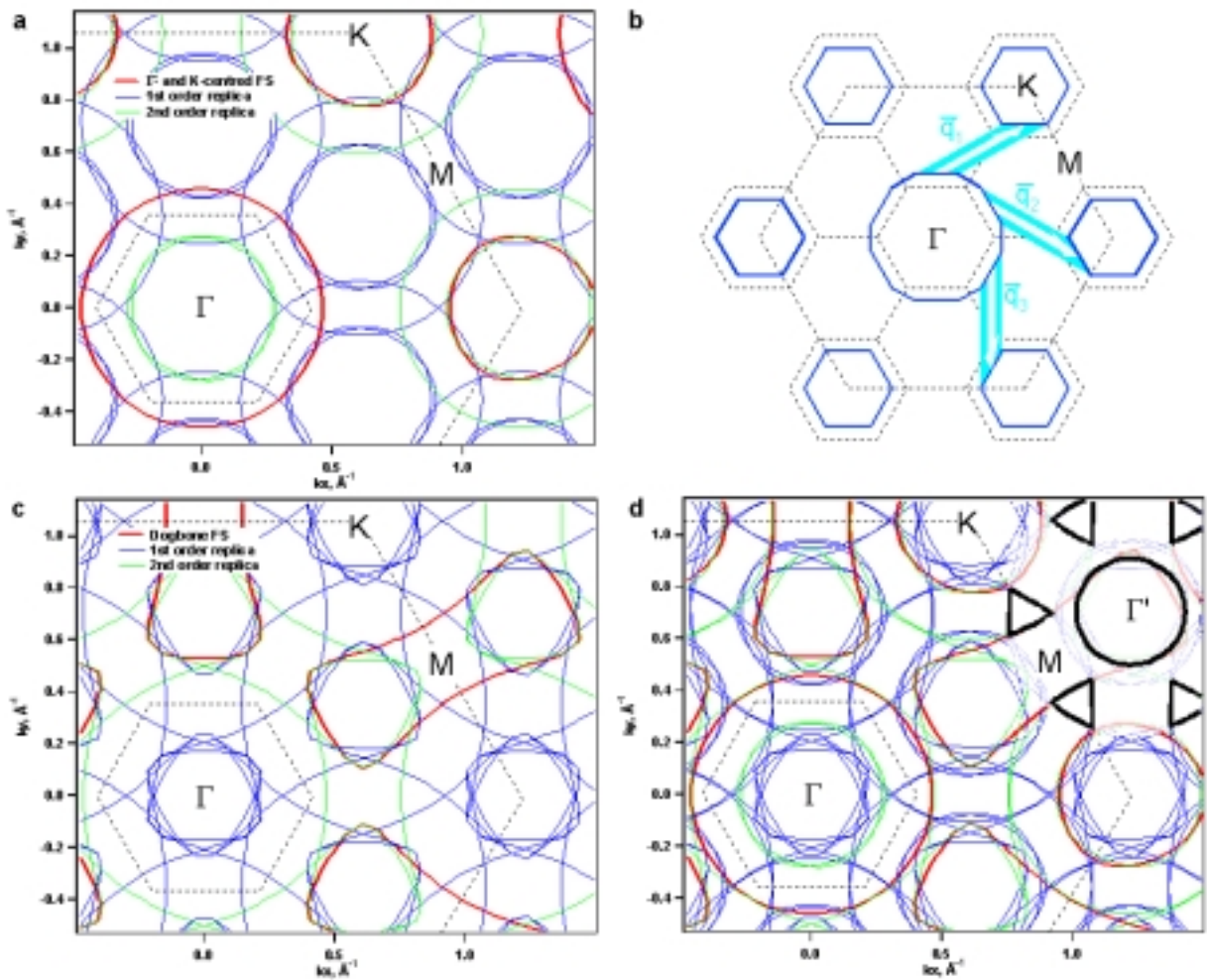

Here we explain in more details the nesting properties of the normal state FS and how to understand the 3x3 folding upon entering the commensurate CDW regime.

In panel **a** we show only Γ- and K-centred barrels and their first order ($\pm \vec{q}_1$, $\pm \vec{q}_2$, $\pm \vec{q}_3$) and second order ($\vec{q}_1 + \vec{q}_2$, $\vec{q}_2 + \vec{q}_3$) replica. One can see that the nesting conditions are nearly perfect for all points of the K-barrel: shifting the Γ-barrel by one of the CDW vectors results in an overlap with the corresponding part of the K-barrel with opposite Fermi velocities. Panel **b** shows an idealized situation, which is actually very close to the real one, since a more careful examination of Fig. 1a reveals the straight sections of both Γ- and K-barrels.

This is, however, not enough to explain the fully gapped K-barrel in the CDW state.

In panel **a** it is seen that 3x3 folding produces a double-walled barrels in the centre of the new BZ which are supposed to interact with the single copies of Γ-barrel. This is not surprising as the original BZ contains only one complete Γ-barrel and two complete K-barrels. It means, that the K-barrels of a system with the FS schematically shown in panel **b**, will not disappear after a 3x3 folding.

Panel **c** shows the dogbone FS centred around M-points together with its first- and second order replica. Folding of this FS sheet results in the same triangular FSs as in panel **a**, a rather complicated set of features around the centres of the new BZ and nearly exact copies of the Γ-barrels and its replica. It is these copies of the Γ-barrels which were missing in panel **a** to interact with double-walled K-barrels.

Panel **d**, where all FSs and their folded replica are shown, summarizes all mentioned above. When hybridization effects are switched on, the former K-barrel completely disappears, doubly degenerate triangular FSs emerge around the corners of the new BZ and the complicated set of features seems to evolve into a four-times degenerate nearly circle FSs around the Γ"- points (at least experimentally we can currently resolve only these semi-circular FSs and not two doubly degenerate hexagons with rounded corners as is suggested by the panel **c**).